\documentclass[journal]{IEEEtran}
\ifCLASSINFOpdf
  \usepackage[pdftex]{graphicx}
\else
  \usepackage[dvips]{graphicx}
\fi
\usepackage{amsmath}
\usepackage{algorithmic}

\usepackage{array}
\usepackage{url}

\usepackage{booktabs}
\usepackage{xcolor}
\usepackage{amssymb}
\usepackage{amsfonts}
\usepackage{algorithm}
\usepackage{tabularx}
\newcolumntype{L}{>{\raggedright\arraybackslash}X}

\newcolumntype{R}{>{\raggedleft\arraybackslash}X}

\newcolumntype{C}{>{\centering\arraybackslash}X}
\usepackage{soul}
\usepackage{xcolor}

\sethlcolor{yellow}

\begin{document}
%
\title{Hardware-Accelerated Algorithm for Complex Function Roots Density Graph Plotting}
%
%
%

\author{Ruibai~Tang,~\IEEEmembership{Graduate Student Member,~IEEE,}
        and~Chengbin~Quan,~\IEEEmembership{Senior Member,~IEEE}
\thanks{R. Tang is with the Zhili College, Tsinghua University, Beijing 100084, China. E-mail: \texttt{trb21@tsinghua.org.cn}.}
\thanks{C. Quan is with the Department of Computer Science and Technology, Tsinghua University, Beijing 100084, China. E-mail: \texttt{quancb@tsinghua.edu.cn}}
}

\maketitle

\begin{abstract}
Solving and visualizing the potential roots of complex functions is essential in both theoretical and applied domains, yet often computationally intensive. We present a hardware-accelerated algorithm for complex function roots density graph plotting by approximating functions with polynomials and solving their roots using single-shift QR iteration. By leveraging the Hessenberg structure of companion matrices and optimizing QR decomposition with Givens rotations, we design a pipelined FPGA architecture capable of processing a large amount of polynomials with high throughput. Our implementation achieves up to 65× higher energy efficiency than CPU-based approaches, and while it trails modern GPUs in performance. Compared with state-of-the-art QR decomposition solutions,
our design specificly optimize QR decomposition for complex-valued Hessenberg matrices up to size 6x6, exhibiting a moderate throughput of 16.5M QR decompositions per second, while prior works
have predominantly focused on 4x4 general matrices. 
\end{abstract}


%
\IEEEpeerreviewmaketitle

\section{Introduction}
%
%
%
%
\IEEEPARstart{I}{n} mathematical analysis, investigating the properties of solutions to equations—particularly the existence of such solutions—is a fundamental and central problem. Whether in the context of pure mathematical theory or in the construction of models in physics, engineering, and applied sciences, the validity of many theoretical inferences critically depends on the existence of solutions to equations. 

In the observation stage of equation solutions, researchers frequently resort to numerical methods by discretizing the problem and applying computational algorithms to approximate the distribution of solutions over the domain. However, numerical approaches typically face two key challenges. \textbf{First}, as the discretization becomes finer, the number of sampling points increases dramatically, leading to large-scale scientific computations that are often time-consuming. \textbf{Second}, the numerical process may generate a vast amount of raw output data, which requires additional processing and analysis to extract meaningful insights or to inspire further innovative research. Some of these tasks, such as data visualization, are particularly common and labor-intensive.

{
The contributions of our work can be summarized as follow:

\begin{itemize}
    \item[1)] We implement a novel hardware system to solve batched complex-valued polynomial roots. Our approach is turning to solve the eigenvalues of the Frobenius companion matrix of a complex polynomial in place by leveraging the Hessenberg form and Givens rotation to reduce the storage and the calculation.
    \item[2)] We attach a 1920x1080@60Hz graphic module to our design, so that users can immediately see the dynamic visualization while calculating.
    \item[3)] Our work is open-sourced on Github\footnote{https://github.com/trrbivial/HPC-Auxiliary-Plotter} to facilitate inspection and reuse by readers.
\end{itemize}}

\section{Problem Description}

Consider the following general problem: given a complex function $f(z)$ defined over a subset $D\subset \mathbb{C}$, visualize the distribution of its zeros, i.e., the solutions to $f(z)=0$. A straightforward approach is to partition the domain $D$ into a large number of small subregions $D_i$, and within each region, numerically approximate $f$ using a sequence of polynomial basis to obtain an approximate polynomial $g$, such that $||f-g||< \varepsilon$. If $g$ has a root within $D_i$, then $f$ is likely to have a root in the same region, and the distribution of zeros of $g$ serves as a good approximation to that of $f$. 

{
\subsection{Polynomial approximation of $f(z)$ on the unit disk}

Least-squares ($L^2$) approximation is a common numerical method to approximate a complex-valued target function
$f:\overline{D}\to\mathbb{C}, D=\{z\in\mathbb{C}:|z|<1\}$ by polynomials 
$p_n(z)=\sum_{k=0}^n a_k z^k$ with complex coefficients $a_k$.

Defining the inner product $\langle g,h\rangle = \int_D g(z)\overline{h(z)}{\rm d}z$, and solving the equation \(G\mathbf{a}=\mathbf{b}\), where
\(G_{j,k}=\langle z^j,z^k\rangle\) and 
\(b_j=\langle f,z^j\rangle\), We can get the coefficients $a_k$ of the polynomial $p_{n}(z)$.

In practical scenarios, $n=6$ usually meets the accuracy requirement of $10^{-3}$. Consider a rational function with a pole outside the unit disk, $f(z)=\frac{1}{1-0.3z}$. Using $L^2$ approximation, we verify that  
$||f-p_6||_{\infty} \le 7.3\times 10^{-4}$ on the unit disk. Besides, on the smaller disk, $|z|<\delta =10^{-2}$, the $L^2$ approximation $\hat{p}_6$ can achieve accuracy $||f-\hat{p}_6||_{\infty}\le 10^{-14}$. In most cases, e.g. the function is locally continuous, we can reduce the size of the disc to achieve significant accuracy improvement, and the smaller disk size can always be selected when the numerical sampling gets finer. Therefore, we choose $n\le6$ to implement our design. }

\subsection{Solving the Roots of a Polynomial}

For a complex polynomial $P(z)=z^{n}+a_{n-1}z^{n-1}+...+a_0$, according to linear algebra knowledge, the eigenvalues of its \textbf{Frobenius} companion matrix correspond precisely to the roots of the polynomial. Therefore, one can equivalently compute the eigenvalues of its associated companion matrix.

The \textbf{QR iteration with single shift}, as shown in Alg.~\ref{alg:QR_Iter_with_single_shift}, is known for its rapid convergence properties to solve all eigenvalues of a small matrix. 

\begin{algorithm}[h]
\footnotesize
    \caption{QR Iteration with Single Shift}
    \label{alg:QR_Iter_with_single_shift}
    \renewcommand{\algorithmicrequire}{\textbf{Input:}}
    \renewcommand{\algorithmicensure}{\textbf{Output:}}
    
    \begin{algorithmic}[1]
        \REQUIRE $A\in M_n(\mathbb{C})$ ($a_{i,j},0\le i, j\le n-1$) 
        \ENSURE $eig[0...n-1]$ 
            \STATE $m \gets n$
            \WHILE{$m \ge 2$}
                \FOR{$i = 0,1, ..., \texttt{iter} $}
                    \STATE $s \gets a_{m-1,m-1}$
                    \STATE $A \gets A - sI_m$
                    
                    \STATE $(Q, R) \gets \operatorname{QR\_DECOMP}(A)$ 
                    \STATE $A \gets RQ + sI_m$
                \ENDFOR
                \STATE $eig[m-1] \gets a_{m-1,m-1}$
                \STATE $m \gets m - 1$
                \STATE $A \gets A[0...m - 1][0...m - 1]$
            \ENDWHILE
            \STATE $eig[0] \gets a_{0,0}$
            \RETURN $eig[0...n-1]$
    \end{algorithmic}
\end{algorithm}

\subsection{QR Decomposition using Givens Rotation}

One effective method for performing QR decomposition is the \textbf{Givens rotation} \cite{golub2013matrix}, which applies a sequence of plane rotations to zero out selected elements. Each Givens rotation has the following form:
\begin{equation*}
\begin{pmatrix}
c & s \\
-\bar{s} & \bar{c} 
\end{pmatrix}
\begin{pmatrix}
a \\
b  
\end{pmatrix}
=
\begin{pmatrix}
\sqrt{|a|^2 + |b|^2} \\
0  
\end{pmatrix}
\end{equation*}

In our problem, the companion matrix $A$ is a \textbf{Hessenberg matrix}. It can be proved (see Appendix) that throughout the execution of Alg.~\ref{alg:QR_Iter_with_single_shift}, the matrix $A$ preserves its Hessenberg structure at every iteration step. This structural invariance allows for a more compact storage representation and the use of simpler algorithms for matrix operations.

Let $Q_i$ denote the unitary matrix representing the Givens rotation applied at step $i$, where $i =1,2,...,n-1$.
\begin{equation*}
Q_i = \begin{pmatrix}
I_{i-1} & & & \\
& c_i & s_i &   \\
& -\bar{s_i} & \bar{c_i} &   \\
& &  & I_{n-1-i}  
\end{pmatrix}
\end{equation*}

Each iteration in the Algorithm (\ref{alg:QR_Iter_with_single_shift}) includes the following step: $A \gets RQ = (Q_{n-1}...Q_2Q_1)A(Q_1^HQ_2^H...Q_{n-1}^H)$. As a result, the QR decomposition does not need to be explicitly performed. The matrix $A$ can be updated directly by accumulating the Givens rotation in-place, which means the operation can be dived into two procedures $A \leftarrow (Q_{n-1}...Q_2Q_1)A$ and $A \leftarrow A(Q_1^HQ_2^H...Q_{n-1}^H)$. During the first procedure, every time performing $A\leftarrow Q_iA$, we only need to update \textbf{two rows} of matrix $A$. During the second procedure, every time performing $A \leftarrow AQ_i^H$, similarly, we only need to update \textbf{two columns} of matrix $A$. 

\section{Architecture of Our Design}

\begin{figure}[h]
  \centering
  \includegraphics[width=\linewidth]{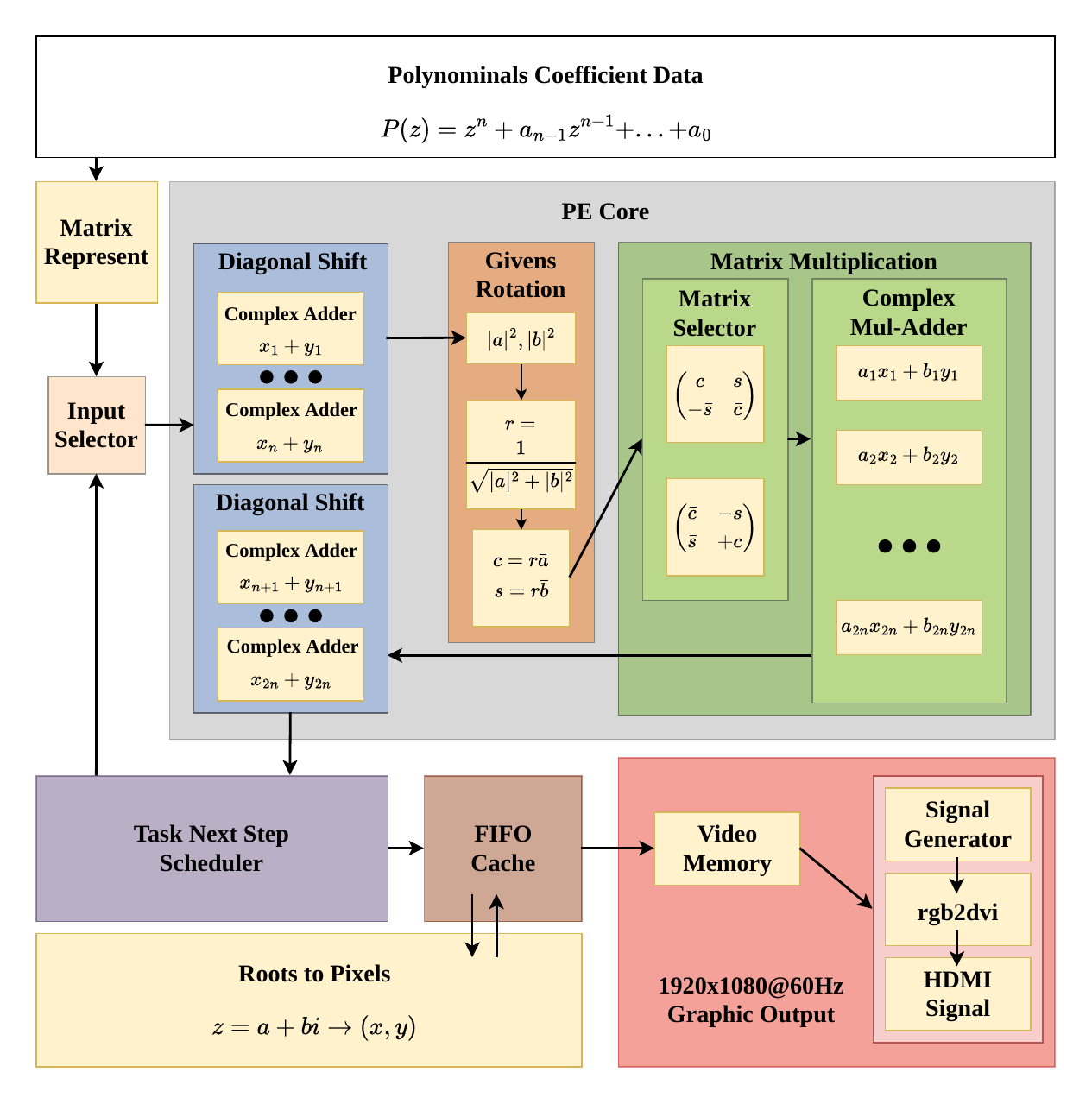}
  \caption{Architecture of Our Design}
  \label{fig:overview_of_ourwork}
\end{figure}

\begin{figure}[h]
  \centering
  \includegraphics[width=\linewidth]{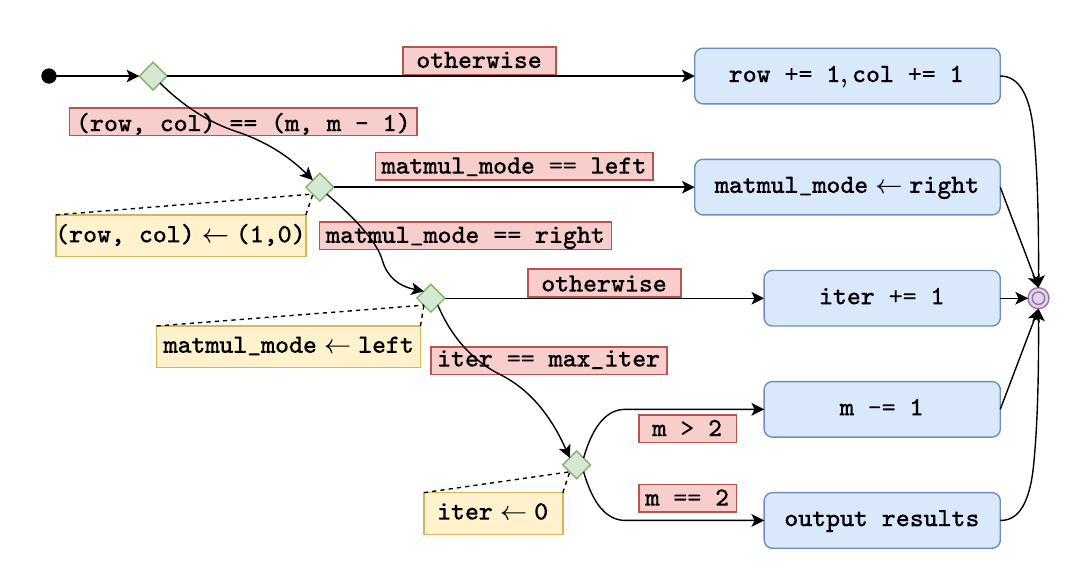}
  \caption{Status Transition Logic in Task Next Step Scheduler}
  \label{fig:status_transition_logic}
\end{figure}

Fig. \ref{fig:overview_of_ourwork} illustrates the architecture of our design. We employ a Task Next Step Scheduler (Fig. \ref{fig:status_transition_logic}) to manage and update the algorithm progress of each matrix. The PE (Processing Element) Core is a non-blocking pipeline. During each pass, one or more internal modules are activated to operate on the matrix, while the inactive modules effectively serve as data buffers. This design eliminates the need for an additional memory pool to store intermediate states.

At the input port of the PE Core, whenever a pipeline gap is detected, the Input Selector prioritizes accepting a task issued from the Task Next Step Scheduler. Otherwise, it fetches a new input from the external source and initiates a new task of Alg.~\ref{alg:QR_Iter_with_single_shift}. If a task satisfies the termination condition after a pass to the PE Core, the eigenvalue results will be written into the FIFO cache within only one cycle to avoid pipeline blocking. Finally, the results stored in FIFO cache will be converted into pixel coordinates on the screen and be written into the video memory for visualizing.

The PE core is composed of four major modules in the pipeline order. The Diagonal Shift modules are triggered only when the input matrix is at the beginning or end of a QR iteration; they add or subtract the shift value $s=a_{m-1,m-1}$ to the diagonal values of the matrix. The Givens Rotation module calculates a pair of rotation parameters $(c,\, s)$ based on matrix entries located at positions $(i,\, j)$ and $(i-1,\, j)$. The Matrix Multiplication module contains $2n$ submodules of $ax + by$.

\begin{table*}[t]
  \centering
  \footnotesize
  \caption{Resource Usage of Modules (Routing Fabric Not Included)}
  \label{tab:resource_utilization}
  \begin{tabular*}{\textwidth}{@{\extracolsep{\fill}}lrrrrr@{}}
    \hline
    \textbf{Module} & \textbf{LUTs} & \textbf{Regs} & \textbf{BRAM} & \textbf{DSPs} & \textbf{Area (mm$^2$)} \\
    \hline
    Total Usage & 48{,}859 & 74{,}283 & 268 & 352 & 44.47 \\
    \hline
    PE Core & 40{,}886 (84\%) & 62{,}796 (85\%) & -- & 316 (90\%) & 24.25 \\
    \quad Diag. Shift (Sub) & 3{,}514 (7\%) & 5{,}482 (7\%) & -- & 12 (3\%) & 1.79 \\
    \quad Givens Rotation & 3{,}996 (8\%) & 4{,}102 (6\%) & -- & 28 (8\%) & 2.12 \\
    \quad \textbf{Matrix Mult.} & \textbf{29{,}712 (61\%)} & \textbf{47{,}992 (65\%)} & -- & \textbf{264 (75\%)} & \textbf{18.53} \\
    \quad Diag. Shift (Add) & 3{,}664 (7\%) & 5{,}220 (7\%) & -- & 12 (3\%) & 1.81 \\
    \hline
    Roots to Pixels & 6{,}655 (14\%) & 10{,}080 (14\%) & -- & 36 (10\%) & 3.62 \\
    Video Memory & 468 (1\%) & 14 (0\%) & \textbf{256 (96\%)} & -- & \textbf{15.50} \\
    FIFO Cache & 287 (1\%) & 214 (0\%) & 12 (4\%) & -- & 0.82 \\
    Others & 563 (1\%) & 1{,}179 (2\%) & -- & -- & 0.28 \\
    \hline
  \end{tabular*}
\end{table*}

\begin{table*}[t]
  \centering
  \footnotesize
  \caption{Comparison of FPGA QR Decomposition Implementations}
  \label{tab:comparison_among_qr_decomp}
  \begin{tabular*}{\textwidth}{@{\extracolsep{\fill}}l|c|c|c|c|c@{}}
    \hline
    \textbf{Work} & \textbf{\cite{7387603}} & \textbf{\cite{7110554}} & \textbf{\cite{6292059}} & \textbf{\cite{7818557}} & \textbf{Ours} \\
    \hline
Algorithm & Sorted QR Decomp & Givens Rotation & Givens Rotation & Modified Gram-Schmidt & Givens Rotation \\
    FPGA Series & Virtex-6 & Virtex-5 & Virtex-5 & Virtex-5 & Artix-7 \\
    Matrix & $4\times4$ (Complex) & $4\times4$ (Real) & $4\times4$ (Real) & $4\times4$ (Complex) & $\le6\times6$ (Complex) \\
    Matrix Type & General & General & General & General & Hessenberg \\
    Precision & 16bit (Fixed) & 16bit (Fixed) & 16bit (Fixed) & 16bit (Fixed) & 32bit (Float)\\
    Frequency (MHz) & 192 & 254 & 246 & 132.80 & 100 \\
    Slice Registers & 24,476 & 2,085 & 16,929 & 6,173 & 62,796 \\
    Slice LUTs & 12,034 & 2,671 & 10,899 & 4,484 & 40,886 \\
    DSPs & 82 & 12 & 28 & 0 & 316 \\
    Throughput ($\times 10^{6}$ QR Decomp/s) & 32 & 31.7 & 1.36 & 44.27 & 16.5 \\
    \hline
  \end{tabular*}
\end{table*}

Finally, there for sure can be multiple PE Cores for parallel computing as long as the FIFO cache has enough capability to handle the stream of the results output. In our experiments, we only employ one PE core and one bus for the FIFO cache to write the video memory.

\section{Evaluation}

\subsection{Pipeline Efficiency Analysis}

Let the original size of the input matrix be $n$, and the current size be $m$, the same as in the Alg.~\ref{alg:QR_Iter_with_single_shift}. Each matrix size level $m$ undergoes QR iterations of $T$ times. Each iteration requires $2m - 2$ passes through the pipeline. The average cycles per input can be calculated by
$$
C_{\rm ave} = T \sum_{m = 2}^{n} (2m - 2) = n(n - 1)T
$$

\subsection{Implementation}

We Implemented the hardware design using \texttt{SystemVerilog} and carried out simulation, synthesis, and implementation using \texttt{Xilinx Vivado 2019.2} \cite{vivado}. The target FPGA device is \texttt{XC7A200T-2FBG484I} \cite{yuanshen}. Detailed resource utilization for each module is shown in Table \ref{tab:resource_utilization}. Area estimations without routing fabric are based on the parameters: LUT(291$\mu m^2$), Reg/FF(96$\mu m^2$), BRAM(0.06$mm^2$), DSP(0.02$mm^2$). An overview of the design is presented in Table \ref{tab:impl_exp_settings}. 

In our implementation, we only employ one PE, but parallelization is an option. The FIFO Cache is the final throughput limit. If only one PE is running, The FIFO Cache operates for only $2\%$ of the clock cycles. There is no problem for the FIFO Cache to handle the outputs from no more than 16 PEs by conservative estimation.

\begin{table}[t]
  \centering
  \footnotesize
  \caption{Implementation and Experiment Settings}
  \label{tab:impl_exp_settings}
  \begin{tabular*}{\columnwidth}{@{\extracolsep{\fill}}ll@{}}
    \hline
    \textbf{Item} & \textbf{Specification} \\
    \hline
    Platform & Xilinx Vivado 2019.2 \\
    FPGA / Tech. & XC7A200T-2FBG484I / 28\,nm \\
    Freq. (PE/VGA) & 100 / 148.5\,MHz \\
    Precision & FP32 \\
    Poly Deg. / QR Iters & $n{=}6$ / $T{=}10$ \\
    Power (PE / Total) & 1.43 / 2.22\,W \\
    Perf. (Avg / Peak) & 13.93 / 20.40\,GFLOP/s \\
    Energy Eff. & 9.74\,GFLOP/(s$\cdot$W) \\
    Parallel Optional? & Yes (Employ $\ge$2\,PEs) \\
    \hline
    OS & Arch Linux x86\_64 {(6.17.8-arch1-1)} \\
    CPU & Intel i7-11800H (16) @ 4.6\,GHz \\
    GPU & NVIDIA RTX 3060 Mobile / Max-Q \\
    \texttt{g++} {(15.2.1)} &  \texttt{-Ofast -fopenmp}\\
    \texttt{nvcc} {(13.0.88)} &  \texttt{-O3 -arch=sm\_80 --use\_fast\_math} \\
    \hline
  \end{tabular*}
\end{table}

\begin{table}[t]
  \centering
  \footnotesize
  \caption{Experiment Results}
  \label{tab:experiment_results}
  \begin{tabular*}{\columnwidth}{@{\extracolsep{\fill}}lrrr@{}}
    \hline
    \textbf{Device} & \textbf{Throughput (/s)} & \textbf{GFLOP/s} & \textbf{GFLOP/(s$\cdot$W)} \\
    \hline
    CPU  & {$1.22{\times}10^5$} & {5.11} & {0.15} \\
    GPU  & {$2.79{\times}10^7$} & {2079.88} & {29.71} \\
    FPGA & $3.33{\times}10^5$ & 13.93 & 9.74 \\
    \hline
  \end{tabular*}
\end{table}

{
\begin{algorithm}[h]
\footnotesize
    \caption{Optimized Algorithm \ref{alg:QR_Iter_with_single_shift} on CPU/GPU}
    \label{alg:QR_CPU_GPU}
    \renewcommand{\algorithmicrequire}{\textbf{Input:}}
    \renewcommand{\algorithmicensure}{\textbf{Output:}}
    
    \begin{algorithmic}[1]
        \REQUIRE $A\in M_n(\mathbb{C})$ ($a_{i,j},0\le i, j\le n-1$) 
        \ENSURE $eig[0...n-1]$ 
            \STATE $\texttt{t} \gets (n-1)*\texttt{iter}$
                \FOR{$i = 0,1, ..., \texttt{t}-1 $}
                    \STATE $m \gets n-i/\texttt{iter};\, d \gets A_{m-1,m-1};\, A \gets A - dI_m$
                    \FOR{$r=1,...,m-1$}
                        \STATE $(c_{r},s_{r}) \gets \operatorname{GIVENS\_COEF}(A_{r-1,r-1}, A_{r,r-1})$
                        \FOR {$j =0,...,m-1$}
                        \STATE $(a,b) \gets (A_{r-1,j},A_{r,j})$
                        \STATE $(A_{r-1,j},A_{r,j})\gets (c_ra+s_rb, -\bar{s_r}a+\bar{c_r}b)$
                        \ENDFOR
                    \ENDFOR
                    \FOR{$r=1,...,m-1$}
                        \FOR {$j =0,...,m-1$}
                        \STATE $(a,b) \gets (A_{j, r-1},A_{j, r})$
                        \STATE $(A_{j, r-1},A_{j,r})\gets (\bar{c_r}a+\bar{s_r}b, -s_ra+c_rb)$
                        \ENDFOR
                    \ENDFOR
                    \STATE $A \gets A + dI_m;\, eig[m-1] \gets A_{m-1,m-1}$
                \ENDFOR
            \STATE $eig[0] \gets a_{0,0}$
            \RETURN $eig[0...n-1]$
    \end{algorithmic}
\end{algorithm}}

{
\subsection{Experiment Settings}

We reviewed several representative works. \textbf{(1)} \texttt{cuSOLVER} \cite{nvidia-cusolver} provides the routine \texttt{cusolverDnXgeev} for hybrid CPU/GPU eigenvalue computation, but its batched mode supports only Hermitian matrices. For our non-Hermitian case, we must launch multiple matrices via CUDA streams, achieving only $4.7\times10^4$ matrices/s—far below our custom GPU implementation. Similarly, work \cite{6885312} accelerates for one single large matrix on GPU, but fails to scale for many small ones. Our application instead requires concurrent processing of a large number of small matrices. \textbf{(2)} Works \cite{7387603,7110554,6292059,7818557} represent state-of-the-art FPGA QR decomposition designs, all explicitly computing $Q$ and $R$. In contrast, our architecture fuses operations to perform an implicit QR process, avoiding explicit storage of $Q$, $R$, and fusing in-place $R\times Q$ updates. This fundamental difference makes it challenging to establish a fair performance comparison between our work and explicit QR-based implementations. But we can still compare our work with them by reporting throughput in terms of equivalent implicit QR decompositions per second.

{
We implemented Alg.~\ref{alg:QR_CPU_GPU} using C++ and CUDA, open-sourced on repo\footnote{https://github.com/trrbivial/HPC-Auxiliary-Plotter}. For CPU, we use OpenMP (\texttt{\#pragma omp parallel for}). For GPU, each CUDA block consists of 32 threads and processes four matrices. All matrices assigned to a block share a single shared memory buffer of approximately 1.54KB, containing the four input matrices and their associated Givens rotation coefficients. We assign six consecutive threads to each matrix (thread 0–5, 6–11, 12–17, 18–23), while others (thread 24-25, 26-27, 28-29, 30-31) write the givens coefficients of four matrices into shared memory to cover the memory latency. Each warp can simultaneously process one diagonal, one row, or one column of each matrix. We do not measure the \texttt{memcpy} time for input and output data to be transmitted between host and device. The \texttt{Nsight Compute} shows our GPU implementation has 79.17\% Compute Throughput, 94.16\% Memory Throughput and 33.28\% Occupancy. The measured runtime power is 34.6\,W and 70\,W for CPU and GPU, respectively.
}

Table \ref{tab:impl_exp_settings} summarizes the environment configuration used for experiments, including the operating system, CPU and GPU models, compiler versions, and compilation flags.

\subsection{Results and Comparison}

Table \ref{tab:experiment_results} shows that our design outperforms the CPU in both throughput and energy efficiency, achieving 65× higher energy efficiency. Our implementation delivers about one-third the energy efficiency of the GPU.

Table \ref{tab:comparison_among_qr_decomp} summarizes the performance and resource utilization of our QR decomposition part in comparison with representative state-of-the-art FPGA-based solutions.

\section{Related Works}

\textbf{QR Decomposition.} There are plenty of hardware designs for solving general QR decomposition problem, such as \cite{10.1145/3174243.3174273, 10.1145/1596532.1596535, 8729615, 7856841}. Works \cite{7387603, 7110554, 6292059, 7818557} are specialized QR decomposition for $4\times 4$ small matrices. 

\textbf{Eigenvalues of Matrices.} \cite{1657028, 10.1007/s00034-022-02180-7, 8988376, householder} are hardware approaches to calculate eigenvalues of general matrices. \cite{6885312} is a CUDA approach which uses QR method. 

\textbf{Givens Rotation.} \cite{doi:10.1137/0912042} is a square root and division free Givens rotation hardware design for solving least squares problems. \cite{POCZEKAJLO2025102567} evaluates some new CORDIC algorithms implemented on FPGA for the Givens rotator.

\textbf{Roots of Polynomials.} \cite{10.5555/1734797.1734801, 5442817} both compared their hardware and software implementations for solving real polynomial roots using Newton's method.

\section{Conclusion}

In this paper, we presented a hardware-accelerated algorithm for visualizing the root density distribution of complex functions by approximating them with polynomials and solving their roots via single-shift QR iteration. By leveraging the Hessenberg form of companion matrices and optimizing QR decomposition with Givens rotations, we designed a fully pipelined, resource-efficient hardware architecture implemented on FPGA.

Our system is capable of processing a large amount of polynomials with high throughput. We demonstrated that our FPGA implementation achieves significant performance and energy efficiency gains over CPU-based approaches, reaching a 65× improvement in energy efficiency. However, our implementation lags behind modern GPUs in performance.

The proposed design provides a scalable and open-source solution for complex root visualization, and may serve as a reference for future research in accelerating numerical algorithms via hardware, especially in domains where large-scale root-finding and visualization are crucial.

\ifCLASSOPTIONcaptionsoff
  \newpage
\fi



%

\bibliographystyle{IEEEtran}
\bibliography{bibtex/bib/sample-base}

\appendices
{
\section{Why Hessenberg Matrix Preserves Properties}
At first $A$ is Hessenberg matrix, we need to prove the following: 

(1) $(Q_i...Q_2Q_1)A$ is Hessenberg $\forall i=1,...,n-1$. 

(2) $(Q_{n-1}...Q_2Q_1)A (Q_1^HQ_2^H...Q_{j}^H)$ is Hessenberg $\forall j =1,...,n-1$.

Let $ G_i :=\begin{pmatrix}
c_i & s_i   \\
-\bar{s_i} & \bar{c_i}    \\ 
\end{pmatrix}, F_i :=\begin{pmatrix}
R_{i} & *   \\
 & H_{n-i}    \\ 
\end{pmatrix}$, where $R_m$ is an m-order upper triangular matrix, and $H_m$ is an m-order Hessenberg matrix. By definition, $F_i$ is Hessenberg. According to the block matrix multiplication, we have
\begin{align*}
Q_iF_{i-1}=& \begin{pmatrix}
I_{i-1} & & \\
& G_i &   \\
&  & I_{n-1-i}  
\end{pmatrix}
\begin{pmatrix}
R_{i-1} & * \\
& H_{n-i+1}
\end{pmatrix} \\
= &
\begin{pmatrix}
R_{i-1} & * & * \\
& x & * \\
& & H_{n-i}
\end{pmatrix} =
\begin{pmatrix}
R_{i}  & * \\
& H_{n-i}
\end{pmatrix} \\= &F_i, \,\, \forall i = 1,...,n-1
\end{align*}

Let $F_0 = A$, then $F_i = (Q_i...Q_2Q_1)A$ is Hessenberg.

The second situation is a little bit complicated, let 
\begin{align*}
\hat{F}_i :=
    \begin{pmatrix}
        \delta^T & * & * & * \\
        R_{i-1} & \alpha & \beta & * \\
        & & x & \gamma^T \\
        & & & R_{n-1-i}
    \end{pmatrix} = \begin{pmatrix}
        H_{i} & *  \\
        & R_{n-i}
    \end{pmatrix}
\end{align*} where $\alpha,\beta,\gamma,\delta$ are column vector. By definition, $\hat{F}_i$ is Hessenberg. According to the block matrix multiplication, we have
\begin{align*}
\hat{F}_{i}Q_{i}^H=& 
\begin{pmatrix}
        \delta^T & * & * & * \\
        R_{i-1} & \alpha & \beta & * \\
        & & x & \gamma^T \\
        & & & R_{n-1-i}
    \end{pmatrix}
    \begin{pmatrix}
I_{i-1} & & \\
& G_i^H &   \\
&  & I_{n-1-i}  
\end{pmatrix} \\
= &
\begin{pmatrix}
        \delta^T & * & * & * \\
        R_{i-1} & \alpha' & \beta' & * \\
        & y & x' & \gamma^T \\
        & & & R_{n-1-i}
    \end{pmatrix}  = 
\begin{pmatrix}
H_{i+1}  & * \\
& R_{n-1-i}
\end{pmatrix} \\ = &\hat{F}_{i+1}, \,\, \forall i = 1,...,n-1
\end{align*}

Since $F_{n-1}, \hat{F}_1$ are both upper triangular, let $\hat{F}_1=F_{n-1}$, then $\hat{F}_j =(Q_{n-1}...Q_2Q_1)A(Q_1^HQ_2^H...Q_j^H)$ is Hessenberg.}

%







\end{document}